\documentclass[prb,aps,twocolumn,showkeys]{revtex4-1}

\usepackage{amssymb}
\usepackage{amsmath}
\usepackage{multirow}
\usepackage{booktabs}
\usepackage{epsfig}
\usepackage[version=3]{mhchem}
\usepackage{epstopdf}
\usepackage{graphicx}
\usepackage[sort&compress]{natbib}
\usepackage{color}
\usepackage{soul}
\usepackage{caption}
\usepackage{subfig}
\usepackage{adjustbox}
\usepackage{tabu}
\usepackage{pgf}
\usepackage[paperwidth=210mm,paperheight=297mm,centering,hmargin=2cm,vmargin=2cm]{geometry}
\usepackage[font=scriptsize]{caption}
\begin{document}

\title{General model of phospholipid bilayers in fluid phase within the
single chain mean field theory}
\author{Yachong Guo}
\affiliation {Departament d'Enginyeria Qu\'{\i}mica, Universitat Rovira i Virgili 26 Av.
dels Paisos Catalans, 43007 Tarragona Spain}
\author{Sergey Pogodin}
\affiliation{Institute of Chemical Research of Catalonia, ICIQ, Av. Paisos
Catalans 16, 43007 Tarragona, Spain}
\author{Vladimir A. Baulin$^*$}
\affiliation {Departament d'Enginyeria Qu\'{\i}mica, Universitat Rovira i Virgili 26 Av.
dels Paisos Catalans, 43007 Tarragona Spain}
\date{\today }

\begin{abstract}
Coarse-grained model for saturated (DCPC, DLPC, DMPC, DPPC, DSPC) and
unsaturated (POPC, DOPC) phospholipids is introduced within the Single Chain
Mean Field theory. A single set of parameters adjusted for DMPC bilayers gives an adequate
description of equilibrium and mechanical properties of a range of
saturated lipid molecules that differ only in length of their hydrophobic tails and unsaturated
(POPC, DOPC) phospholipids which have double bonds in the tails. A double bond is modeled with a fixed angle of 120 degrees, while the rest of the parameters are kept the same as saturated lipids. The
thickness of the bilayer and its hydrophobic core, the compressibility and
the equilibrium area per lipid correspond to experimentally measured values
for each lipid, changing linearly with the length of the tail. The model for unsaturated
phospholipids also fetches main thermodynamical properties of the bilayers.
This model is used for an accurate estimation of
the free energies of the compressed or stretched bilayers in stacks or
multilayers and gives reasonable estimates for free energies.
The proposed model may further be used for studies of mixtures
of lipids, small molecule inclusions, interactions of bilayers with embedded
proteins.
\end{abstract}

\maketitle

\section{Introduction}

Cell membranes represent thin and flexible layers separating the interior
of the cells from the environment\cite{alberts_molecular_2008}. Specific
structure of the cell membrane\cite{yeagle_structure_2005} provides the cell
with numerous physiological functions: membranes maintain a stable
metabolism of the intracellular environment, modulate and select small
molecules and bio-molecules that can penetrate inside the cell\cite%
{yeagle_structure_2005}. Functional and structural properties of cell
membranes are strongly related to the structure of lipid molecules\cite%
{dopico2007glance}. Cell membranes are composed of a mixture of different
types of lipids including saturated and unsaturated phospholipids,
cholesterol molecules, fatty acids, proteins and other inclusions\cite%
{yeagle_structure_2005}. Membrane properties and biological functions
provided by cell membranes are ensured by tuned balance of membrane
composition. When this balance is altered, the cell function can be modified
and can even lead to certain diseases\cite{halliwell_role_1990}.

Many functional properties of lipid membranes are determined by collective
phenomena, where many molecules interact with each other and self-assemble
in complex arrangements with internal structure and order\cite{dopico2007glance}. In particular, a detailed microscopic description of
collective phenomena in lipid bilayers may require the study of very large systems comprising of large
number of lipid molecules, where atomistic Molecular Dynamics (MD) simulations are not yet practical\cite{muller_coarse-grained_2003}. Thus, theoretical description of large lipid systems is usually limited to coarse-grained models\cite%
{muller_coarse-grained_2003,muller_biological_2006,venturoli_mesoscopic_2006}%
, where groups of atoms are represented by effective beads, thus reducing number of degrees of freedom in simulated systems. It is
usually accepted that the coarse-graining provides an adequate and
consistent description of equilibrium and structural properties of lipid
bilayers\cite{lyubartsev_multiscale_2005,venturoli_mesoscopic_2006,muller_biological_2006}. Since lipid
molecules are rather short and their conformation space is limited, the
resulted equilibrium structures are determined by amphiphilic structure of
molecule and thus the bilayer composition is not sensitive to details of
coarse-graining\cite{venturoli_mesoscopic_2006}.

Combination of coarse-grained molecular models with mean field theories is the next
step towards description of even larger lipid systems. The mechanical and
structural properties of lipid bilayers can be successfully modeled within
the Single Chain Mean Field (SCMF) theory\cite%
{pogodin_coarse-grained_2010}. The SCMF theory was
originally proposed\cite%
{ben-shaul_chain_1985,szleifer_chain_1985,ben-shaul_chain_1986} to describe the self-assembly of surfactants into
spherical micelles\cite%
{ben-shaul_chain_1985,ben-shaul_chain_1986,al-anber_prediction_2005,gezae_daful_accurate_2011}. Computationally expensive calculation of interactions between molecules is replaced by calculation of interactions of a single molecule in different conformations with a mean field, created by other molecules. In this approach correlations between molecules and fluctuations are neglected, while the output of the theory is equilibrium structures. This theory gives detailed molecular
structure of nanoscale objects self-assembled from relatively short
molecules. Since SCMF theory describes the systems in equilibrium, the free
energy of different self-assembled structures can be obtained directly\cite%
{ben-shaul_chain_1985,szleifer_chain_1985,ben-shaul_chain_1986}, easier than in Molecular Dynamics (MD) and Monte Carlo (MC) simulations. In addition,
modifications of this theory, for example, single chain in mean
field simulations\cite{daoulas_single_2006}, can describe long-wavelength
fluctuations.

In this work we propose a general coarse-grained model for the SCMF theory of most common lipids\cite{mouritsen2005life} found in Nature that have the same polar head and differ only in the length of their aliphatic tails. This model is similar in spirit to 10-beads model described in Ref. \citenum{pogodin_coarse-grained_2010}, but is more accurate in description of equilibrium properties of the bilayers and can be applied to a wider range of lipids. To test the performance of our model we compare the free energy of compressed bilayers with MD simulation results\cite{Smirnova2013} obtained within MARTINI model.

\section{General model of saturated and unsaturated phospholipids}

SCMF theory of lipid bilayers\cite{pogodin_coarse-grained_2010,pogodin_equilibrium_2011} describes lipid molecules in a coarse-grained approximation as a group of connected beads interacting via effective pairwise potentials. Each bead represents several atoms while the number of atoms in the bead depends on the level of coarse-graining. In contrast to widely used Self-Consistent Field (SCF)\cite{fleer_polymers_1993,thompson_benchmarking_2012} theories, there is no \textit{a priori} assumption on the probability distribution of conformations of molecules; instead, representative sampling\cite{ben-shaul_chain_1985} of conformations is generated using Monte Carlo or Rosenbluth methods\cite{rosenbluth_monte_1955}. Thus, this method provides more adequate description for short molecules which have non-Gaussian probability distribution of conformations, which is the case for lipids. Explicit generation of the sampling allows to split the interactions of the molecules into intra- and intermolecular parts. Intra-molecular interactions can be calculated explicitly for each conformation during the generation of the sampling, while intermolecular interactions are replaced by interactions of each conformation with mean-fields created by other molecules or external fields. Thus, the probability of each conformation $\Gamma$ is fixed by a given distribution of mean fields, while the mean fields are calculated as averages over all conformations with their probabilities, hence closing the self-consistency loop. Such strategy allows to write a set of self-consistent non-linear algebraical equations, which can be solved numerically.

SCMF method applied to lipid bilayers is summarized in Ref. \citenum{pogodin_coarse-grained_2010}, here we list the resulting expressions. The probability of each conformation $\Gamma$,
\begin{equation}
\rho(\Gamma)=\frac{1}{Z}e^{-\frac{H_{eff}(\Gamma)}{kT}}
\end{equation}
normalized by the normalization constant $Z$, is given by the effective Hamiltonian, $H_{eff}(\Gamma)$, which is determined by distributions of the mean fields\cite{pogodin_coarse-grained_2010,pogodin_equilibrium_2011},
\begin{align}
\frac{H_{eff}(\Gamma)}{kT}&=U_{0}(\Gamma)+(N-1)\left(\int u_{T}(\Gamma,\mathbf{r})\left\langle c_{T}(\mathbf{r})\right\rangle d\mathbf{r}\right. \notag\\
&+\left.\int u_{H}(\Gamma,\mathbf{r})\left\langle c_{H}(\mathbf{r})\right\rangle d\mathbf{r}\right) \notag\\
&+ \int u_{s}(\Gamma,\mathbf{r})c_{s}(\mathbf{r})d\mathbf{r}-\int\lambda(\mathbf{r})\phi(\Gamma,\mathbf{r})d\mathbf{r}
\end{align}

where $N$ is the number of molecules in the system; $U_{0}(\Gamma)$ is the contribution from intra-molecular interactions of a particular conformation $\Gamma$; $u_T$, $u_H$, $u_s$ and $\phi$ are the contributions of each conformation to the fields of interactions of two types of beads, tails (T), heads (H), solvent molecules (s), and excluded volume, in a particular point in space $\mathbf{r}$. The corresponding mean fields created by the molecules are the averages over all conformations with the corresponding probability $\rho(\Gamma)$ are denoted by angular brackets. Hence,
\begin{eqnarray}
\left\langle c_{T}(\mathbf{r})\right\rangle &=&\int\rho(\Gamma)c_{T}(\Gamma,\mathbf{r})d\Gamma\\
\left\langle c_{H}(\mathbf{r})\right\rangle &=&\int\rho(\Gamma)c_{H}(\Gamma,\mathbf{r})d\Gamma\\
\left\langle \phi(\mathbf{r})\right\rangle &=&\int\rho(\Gamma)\phi(\Gamma,\mathbf{r})d\Gamma\\
v_{s}c_{s}(\mathbf{r})&=&\phi_{0}-N\left\langle \phi(\mathbf{r})\right\rangle
\end{eqnarray}
where $v_s$ is the volume of the solvent bead and $\phi_0$ is the maximum volume fraction, allowed in the system\cite{pogodin_coarse-grained_2010}. These equations are accompanied by the incompressibility condition,

\begin{equation}
v_{s}\lambda(\mathbf{r})=\ln\left(v_{s}c_{s}(\mathbf{r})\right)+N\int\rho(\Gamma)u_{s}(\Gamma,\mathbf{r})d\Gamma
\end{equation}
where $\lambda(\mathbf{r})$ is related to lateral pressure inside the bilayer. The interactions between the beads of lipids and the solvent are described by potential square well, which includes hard-core repulsion at the distances smaller than the sum of radii of the interacting beads and attraction or repulsion within the interaction range. This repulsion between the beads of different molecules is determined by the excluded volume field.

These nonlinear equations give distributions of lipids and solvent molecules at equilibrium and the corresponding total free energy of the system. In addition to the global minimum of the system, the solution of equations may also lead to metastable solutions with higher free energy, which may, in principle, inform on possible metastable states as well as the transition path between stable states.


\begin{figure*}[th]
\centering
\includegraphics[width=16cm]{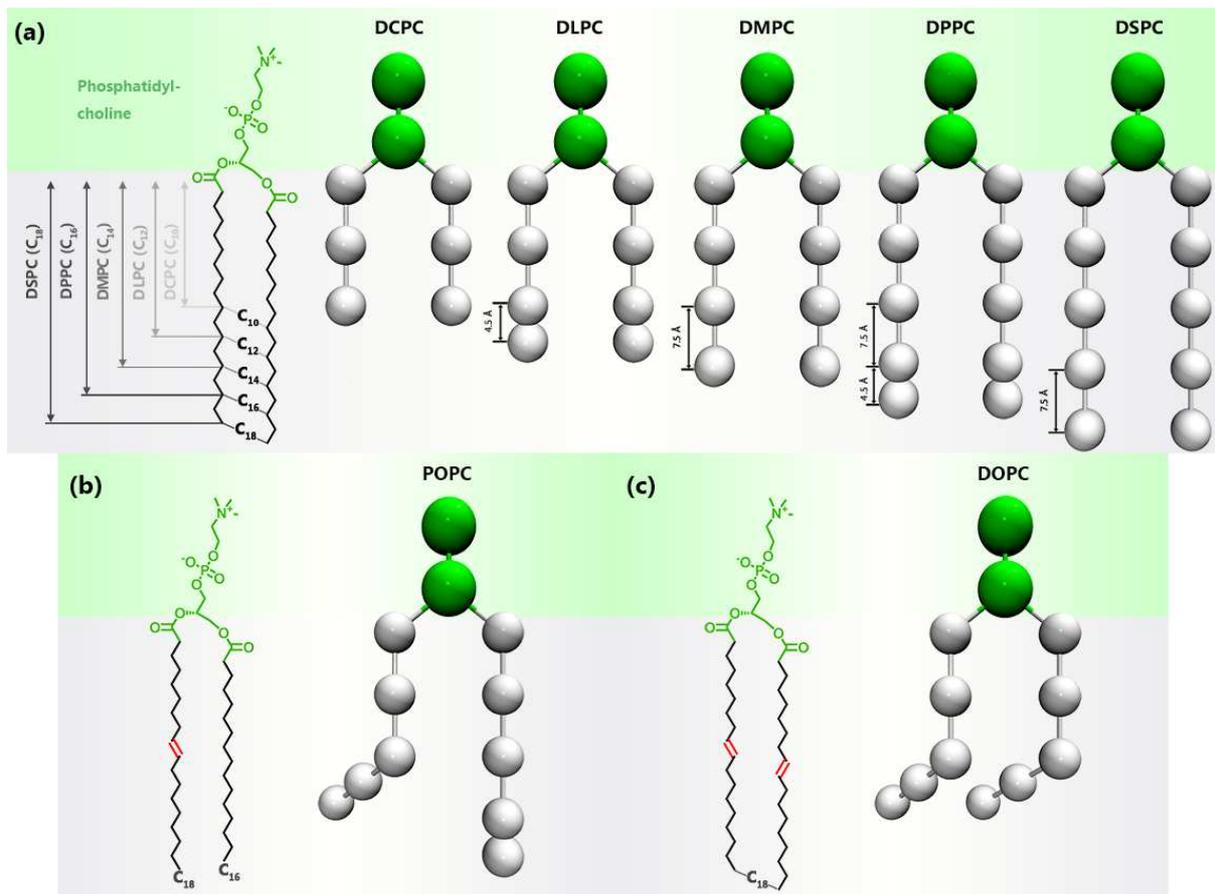}
\caption{(a) Chemical structure (left) and the corresponding coarse grained model (right) of saturated
phospholipids DCPC, DLPC, DMPC, DSPC (from left to right). (b-c) Chemical structure and the coarse grained model of unsaturated phospholipids POPC
and DOPC. Green beads correspond
to hydrophilic (H) and grey beads correspond to hydrophobic (T) groups (see parameters in Table \protect\ref{tab:1}). Unsaturated lipids have a kink of a fixed angle 120 degrees at the double bond.}
\label{fig:saturated}
\end{figure*}


Phospholipid molecules are modeled as a sequence of beads of two types, heads (H) and tails (T), as shown in Figure \ref{fig:saturated}. All studied lipid molecules have the phosphatidylcholine head which is represented by two H-beads of radius equal to 3.3 $\AA$ (Figure \ref{fig:saturated}a), which are the same for all studied lipids.

The saturated lipids differ only in length of the tails of aliphatic chains (Figure \ref{fig:saturated}-a). Thus, the coarse grained model represents the tails of lipids with hydrophobic T-beads of radius 2.5 $\AA$, effectively grouping two or four carbon atoms according to the following rules: (i) the first T-bead, connected to the head, represents one carboxylate and one carbon group; (ii) next non-terminal T-beads represent four carbon groups; (iii) the terminal T-bead of the tail represents two or four carbon groups, depending on the length of the terminal bond (4.5 $\AA$ or 7.5 $\AA$, correspondingly). The length of the rest of bonds is 7.5 $\AA$, and there is no restrictions on the angles between them, \textit{i.e.} the beads are freely jointed. The molecules are generated by self-avoiding walk using Rosenbluth algorithm, i.e. the beads in the resulting conformations do not intersect. The solvent molecules are represented by S-beads of radius 2.5 $\AA$. The parameters of the model are summarized in Table \ref{tab:1}. Two types of beads tails (T) and heads (H) and solvent (S) interact through square well potentials: the interaction is equal to zero if the distance between the center of bead is larger
than the interaction range, while if the centers of beads are within the interaction range, the interaction between the beads has a constant value listed in the table. There are only two types of interactions: between two hydrophobic beads TT and between one hydrophilic bead and solvent HS. Since the molecules are modeled as sequence of hard spheres, close packing is achieved for occupied volume fraction $\phi_0$ smaller than 1, which is, in fact, a parameter of the system controlling the excluded volume interactions. In the model it is set to $0.685$. The parameters of the model are adjusted through series of simulations with large
sampling $(7\times 10^{7})$ and resolution.
The simulation box (Width $\times$ Length $\times$ Height) is set to $100.0$ $\AA$ $\times 100.0$ $\AA$ $\times 62.7$ $\AA$ and the periodic boundary conditions in lateral directions are used.

\begin{table}[th]
\caption{Parameters of the coarse-grain model of saturated and unsaturated phospholipids. Phosholipids of each group
differ only in the length of the hydrophobic tail (Figure \protect\ref%
{fig:saturated}) while the interaction parameters are the same.}
\label{tab:1}{\small \
\begin{tabular}{lcc}
\hline
Lipid model parameters: & \multicolumn{2}{c}{Saturated} \\ \hline
H-bead radius (\AA ) & \multicolumn{2}{c}{$3.3$} \\
T-bead radius (\AA ) & \multicolumn{2}{c}{$2.5$} \\
S-bead radius (\AA ) & \multicolumn{2}{c}{$2.5$} \\
Interaction range (\AA ) & \multicolumn{2}{c}{$7.5$} \\
T-T contact energy (kT) & \multicolumn{2}{c}{$1.20$} \\
H-S contact energy (kT) & \multicolumn{2}{c}{$-0.15$} \\
Bond length (\AA ) & \multicolumn{2}{c}{$7.5$} \\
Terminal group bond length (\AA ) & \multicolumn{2}{c}{$4.5$} \\
Occupied volume fraction $\phi _{0}$ & \multicolumn{2}{c}{$0.685$} \\
&  &  \\ \hline
Simulation parameters: &  &  \\ \hline
Sampling (number of configurations) & \multicolumn{2}{c}{$7\times 10^{7}$}
\\
Simulation box size (\AA ) & \multicolumn{2}{c}{$100.0\times 100.0\times 62.7
$} \\ \hline
\end{tabular}
}
\end{table}

The constructed model for saturated lipids can further be extended to unsaturated lipids of a similar structure, POPC and DOPC, which have exactly the same head group and similar to saturated lipids DPPC and DSPC chemical structure (Figure \ref{fig:saturated}). However, there are important differences in the molecule' structure that should be reflected in the model. (i) One of the tails of POPC is shorter than the other; (ii) POPC molecule has one and DOPC molecule has two double-bonds in the middle of the fatty acid tails. Therefore, tails of POPC have different lengths, but tails of DOPC are of the same length. As a result, in our model (Figure~\ref{fig:saturated}-b,c) the lipid tail with no double bond in POPC has the same structure as the tails of DPPC, while the tail with hydrogen bond in POPC and both tails of DOPC have similar structure as DSPC. In addition, the bead in the middle of the tail corresponding to double bound has a fixed angle of $120$ degrees.

\begin{figure}[!htb]
\centering
\includegraphics[width=8.5cm]{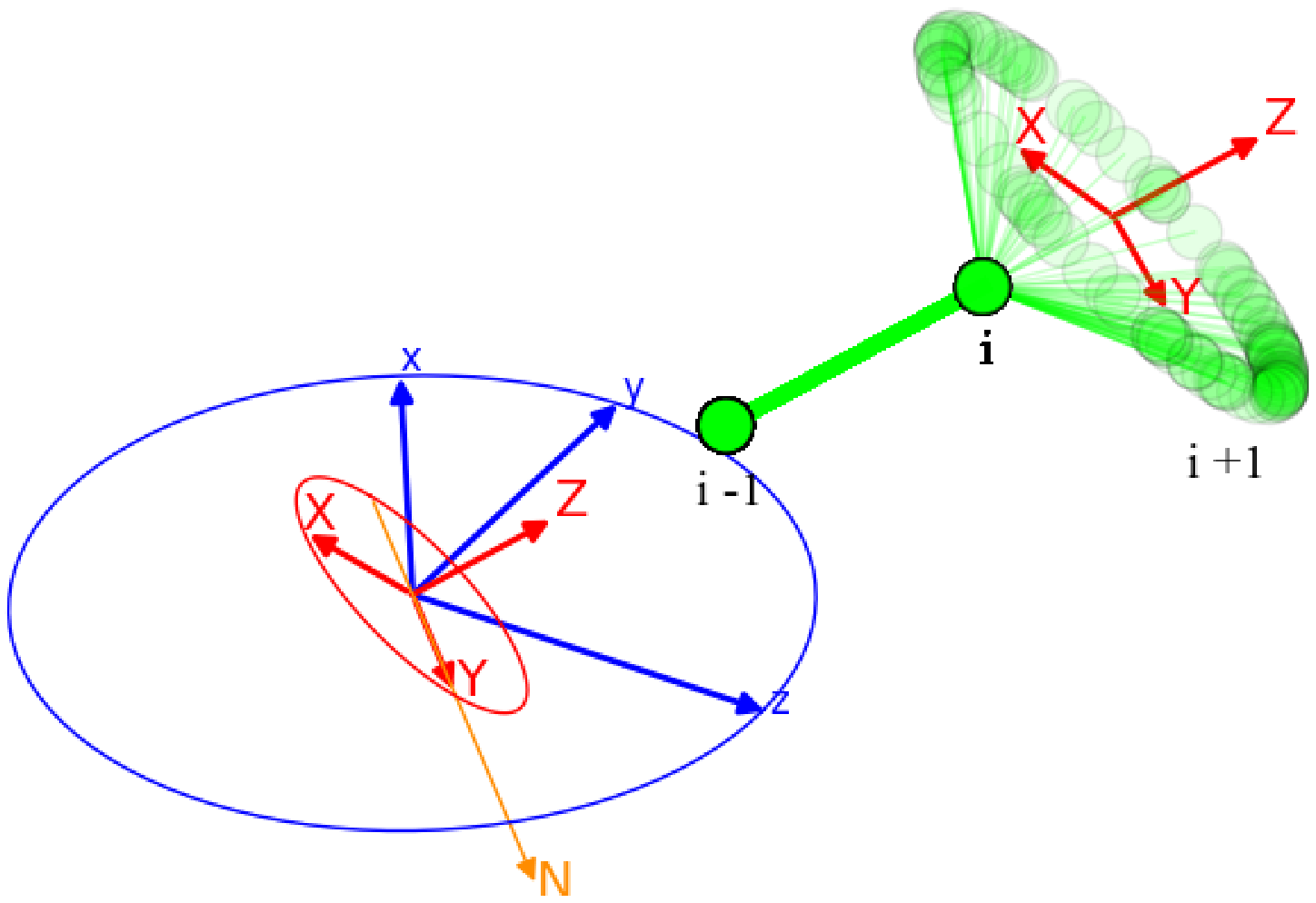}
\caption{Implementation of a fixed angle of a double bond in the unsaturated lipid model.}
\label{fig:angle}
\end{figure}

The implementation of fixed angle is illustrated in Figure~\ref{fig:angle}. A rotated coordinate system $XYZ$ (red) centering on the unsaturated bead $i$ is defined with respect to a fixed coordinate system $xyz$, where $z$-axis is oriented along the extended line of the bond connecting bead $i$ with the bead $i-1$. $xy$ plane is perpendicular to $z$-axis. The line of nodes-$N$ (orange) is defined as the intersection of the $xy$ and $XY$ coordinate planes. $\alpha$ is the angle between $x$-axis and $N$-axis, $\beta$ is the angle between the $z$-axis and the $Z$-axis. Random position of the next bead $i+1$ is generated first in the rotated coordinate system $XYZ$ and than transferred with rotated matrix:
\begin{equation}
\begin{bmatrix}
       \cos \alpha & -\cos \beta \cos \alpha & \sin \beta \sin \alpha           \\
       \sin \alpha & \cos \beta \cos \alpha  & -\sin \beta \cos \alpha \\
       0  & \sin \beta & \cos \beta
\end{bmatrix}
\end{equation}%
to the original coordinate system $xyz$.

\section{Equilibrium properties of lipid bilayers}

The equilibrium properties of unconstrained bilayers at 30 $^\circ$C  are calculated and compared with the experimental data for saturated DCPC, DLPC, DMPC, DPPC and unsaturated POPC, DOPC lipids, see Table \ref{tab:2}. The calculated equilibrium properties include area per lipid, volume per lipid, membrane thickness (defined by the bilayer distance between midpoint of the total volume fraction), hydrophobic core thickness (defined by the bilayer distance between midpoint of the tail beads volume fraction), distance between heads (defined by the peak value of the
slopes of the head beads volume fraction) and the compressibility constant which describes
the rigidity of membrane in lateral direction. It is calculated as a second derivative of the free energy versus area per lipid around the minimum of the free energy (see Ref. \citenum{pogodin_coarse-grained_2010}).

\begin{table*}[htb]
\caption{Equilibrium thermodynamic and mechanical properties of saturated and unsaturated lipids obtained with the SCMF model of lipids and compared with experimental data. Theoretical data for unsaturated lipids corresponds to corrected for volume data denoted by a star in Figure \ref{fig:2}.}
\label{tab:2}
\begin{adjustbox}{center,width=1\textwidth} 
\begin{tabular}{ccccccc!{\vrule width 2pt}cc}
\hline
\multicolumn{1}{c}{\textbf{Lipid}} &  & \textbf{DCPC} & \textbf{DLPC} & \textbf{DMPC} & \textbf{DPPC} & \textbf{DSPC} & \textbf{POPC} & \textbf{DOPC} \\ \hline
\midrule
\multirow{1}[1]{*}{\textbf{Equilibrium Area per lipid }} & Exp   & -     & 63.2 \textsuperscript{\emph{a}} & 60.6 \textsuperscript{\emph{b}} & 59.96 \textsuperscript{\emph{*}}& 58.79\textsuperscript{\emph{*}} & 68.3 & 72.4    \\
      $\mathbf{A_{0}}$  ($\textit{\AA}^{2} $)& Theor & 61.9  & 62.5  & 61.5  & 60.1  & 59   & 65.4  & 69.9 \\
\multirow{1}[0]{*}{\textbf{Volume per lipid }} & Exp   & -     & 991 \textsuperscript{\emph{a}}  & 1101  \textsuperscript{\emph{b}}& -     & -    & 1256  & 1303  \\
       $\mathbf{V_{L}}$ ($\textit{\AA}^{3} $)& Theor & 978   & 1025  & 1098  & 1196  & 1283  & 1306 & 1389\\
\multirow{1}[0]{*}{\textbf{Membrane thickness }} & Exp   & -     & 38.9 \textsuperscript{\emph{a}} & 44.2 \textsuperscript{\emph{b}} & -     & -     & 45.1 & 44.8  \\
       $\mathbf{D_{M}}$ ($\textit{\AA}$)& Theor & 35.7  & 39.7  & 43.8  & 45.7  & 48.3  & 45.8 & 45.2\\
\multirow{1}[0]{*}{\textbf{Hydrophobic core thickness }} & Exp   & -     & 20.9 \textsuperscript{\emph{a}} & 25.4 \textsuperscript{\emph{d}} & 27.3 \textsuperscript{\emph{*}}& -    & 27.1   & 26.8  \\
       $\mathbf{D_{T}}$ ($\textit{\AA}$)& Theor & 19.0  & 23.2  & 25.2  & 27.6  & 30.7  & 27.8 & 27.7\\
\multirow{1}[0]{*}{\textbf{Distance between heads }} & Exp   & -     & 30.8 \textsuperscript{\emph{a}} & 35.3  \textsuperscript{\emph{b}}& -     & -     & 37  & 36.7 \\
       $\mathbf{D_{H}}$ ($\textit{\AA}$)& Theor & 27.3  & 30.9  & 34.8  & 37.5  & 40.1  & 37.8 & 37.5\\
\multirow{1}[1]{*}{\textbf{Compressibility constant }} & Exp   & -     & 302  \textsuperscript{\emph{e}} & 257  \textsuperscript{\emph{e}} & -     & -      & 278\textsuperscript{\emph{e}} & 251 \textsuperscript{\emph{e}}  \\
       \textbf{$\mathbf{K_{C}}$ ($dyn/cm$)}& Theor & 313   & 295   & 272   & 264   & 248   & 292 & 279\\ \hline
\bottomrule
\end{tabular}%
\end{adjustbox}
\par
\textsuperscript{\emph{a}} Experimental data by %
\citeauthor{kucerka_structure_2005}\cite{kucerka_structure_2005} %
\textsuperscript{\emph{b}} Experimental data by %
\citeauthor{kuvcerka2006structure}\cite{kuvcerka2006structure} %
\textsuperscript{\emph{c}} Experimental data by \citeauthor{nagle1996x}\cite%
{nagle1996x}
\textsuperscript{\emph{d}} Experimental data by \citeauthor{feig2008implicit}%
\cite{feig2008implicit}
\textsuperscript{\emph{e}} Experimental data by %
\citeauthor{mathai_structural_2008}\cite{mathai_structural_2008}
\textsuperscript{\emph{*}} DPPC and DSPC are in gel phase at 30 $^\circ$C and these data are extrapolated values by
\citeauthor{kucerka_fluid_2011}\cite{kucerka_fluid_2011} to fluid phase values.
\end{table*}

The experimental data is
collected from different sources of X-ray scattering and corresponds to fully hydrated fluid phase. The temperature of all experimental data is 30 $^\circ$C, which makes it possible the comparison with simulation data. However, the main transition temperature of DPPC and
DSPC is around\cite{gauger_chain-length_2001} 41 $^\circ$C and 54 $^\circ$C correspondingly. This means that these lipid bilayers should be in a gel phase
at 30 $^\circ$C, at which the bilayer has completely different equilibrium and mechanical properties. In order to consider systematically all lipids with different tail lengths in the same framework, it was proposed to use experimental estimates for the effective values that lipid bilayers would have at
30 $^\circ$C if they would not have undergone the transition into gel phase. To do so, the averaged molecular area expansion
 $k_{DPPC} = 0.190$ $\textit{\AA}^{2}/deg$, $k_{DSPC} = 0.167$ $\textit{\AA}^{2}/deg$\cite{kucerka_fluid_2011} was introduced\cite{kucerka_fluid_2011} to extrapolate experimental
 area per lipid to 30 $^\circ$C. Similarly, averaged molecular thickness expansion
 $k_{DPPC} = 0.090$ $\textit{\AA}/deg$ was used to extrapolate the membrane thickness.
However, the compressibility modulus of DPPC is not accessible at such temperature.

\begin{figure*}[htb]
\centering
\includegraphics[width=17cm]{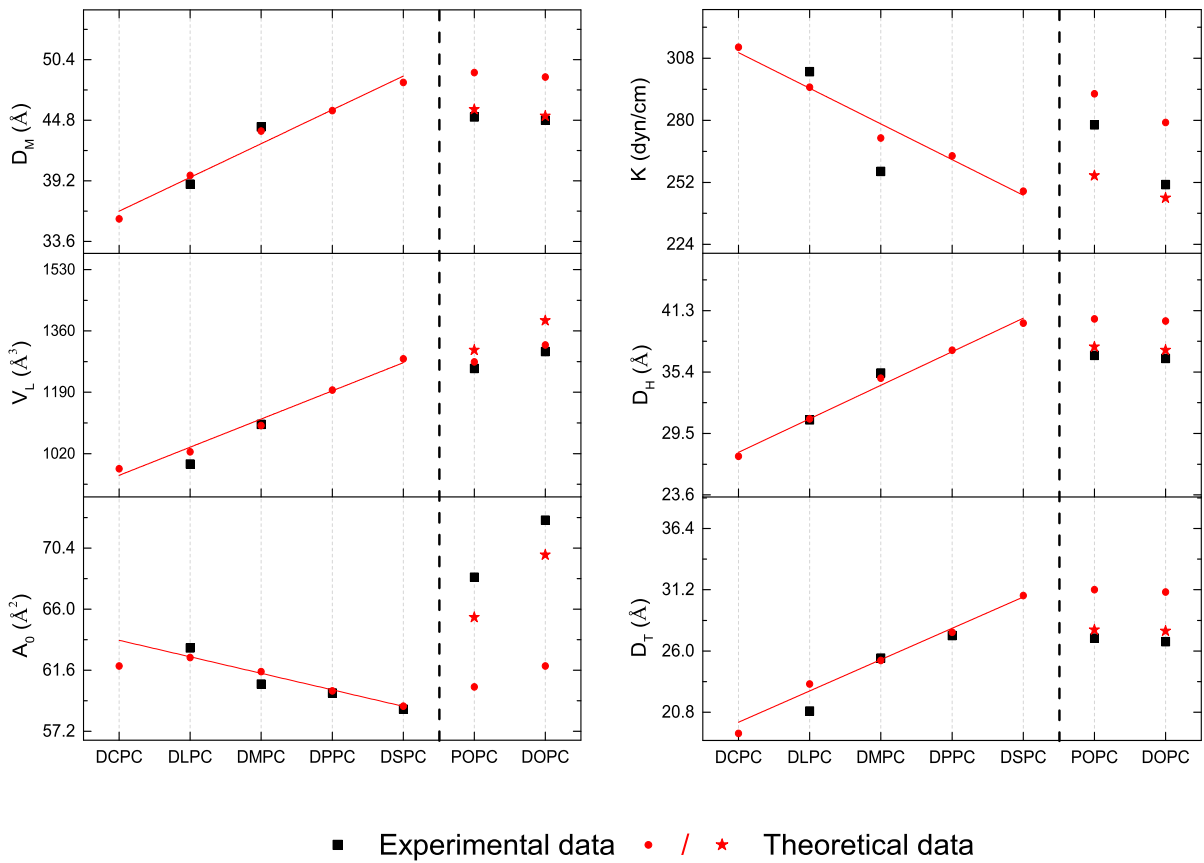}
\caption{Equilibrium thermodynamic and mechanical properties of saturated and unsaturated lipids obtained with the general model of lipids in SCMF theory and compared with experimental data.
solid square represent the experimental data of DCPC, DLPC, DMPC, DPPC, DSPC, POPC, DOPC, Solid sphere reprenents the simulation data obtained with coarsed grained model with SCMF theory, solid star denote the simulation data of POPC, DOPC after considering the effect of excluded volume. A fitting curve is generated as red line based upon the simulation data of saturated lipids}
\label{fig:2}
\end{figure*}

\begin{figure}[t]
\centering
\includegraphics[width=8.5cm]{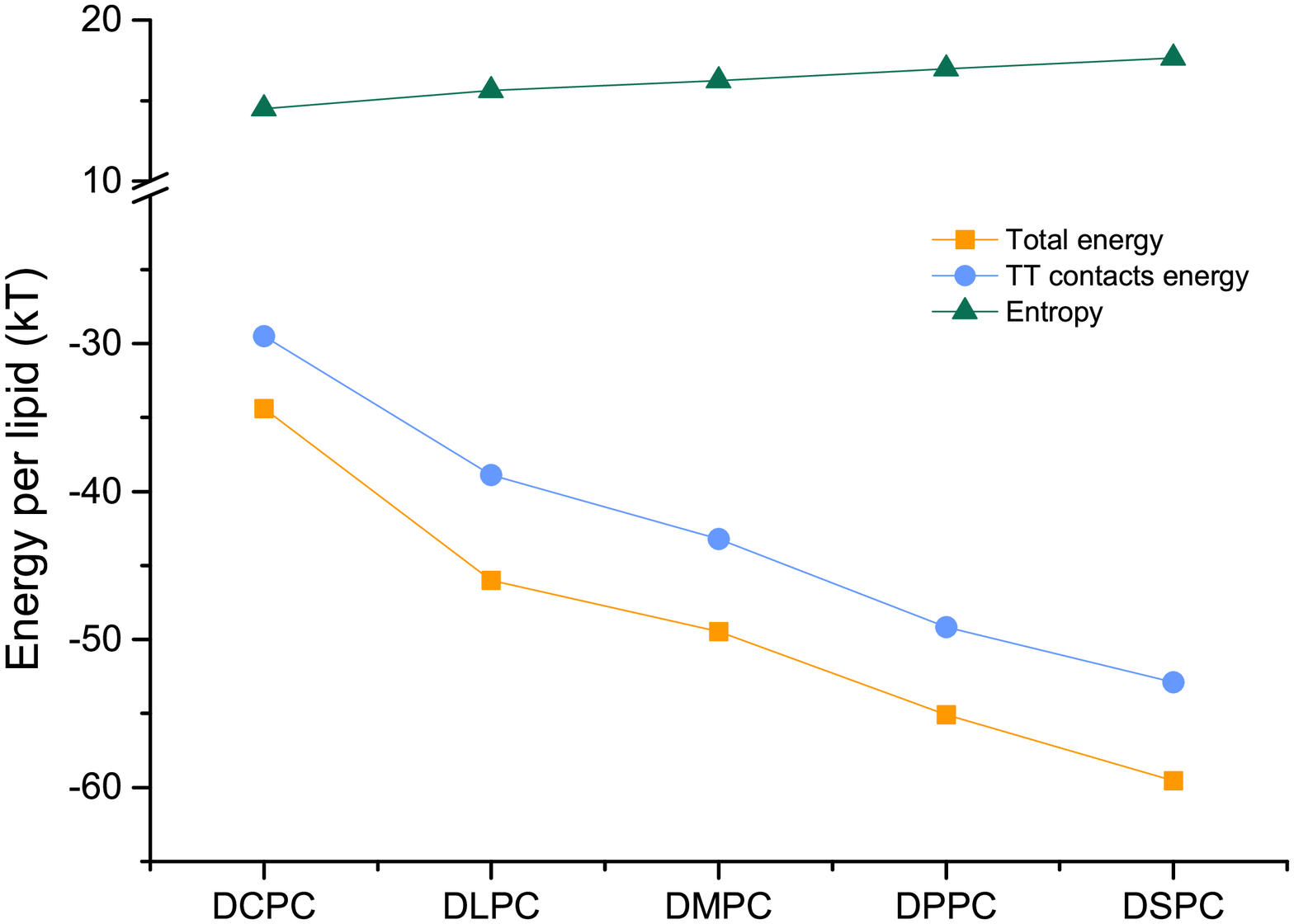}
\caption{Contributions to the total free energy (squares) of equilibrium  saturated fatty acid  liquid PC bilayers calculated within SCMF theory: the energy of TT contacts (sircles) and entropy (triangles). Other contributions (intra-molecular energy and the energy of Hs contacts are small.}
\label{fig:energy}
\end{figure}

\subsection{Saturated lipids}

Good agreement between experimental and theoretical predictions of the general model is shown in Figure \ref{fig:2}. Our model reproduces almost perfectly the equilibrium properties of bilayers assembled from saturated lipids with different tail lengths using the single set of interaction parameters. The calculated equilibrium properties for liquid phase show linear dependance with the length of the tail. The statistical error of SCMF calculations is of the order of 1\% for thickness and area and 5.4\% for compressibility constant since it is a second derivative of the free energy and thus it requires large sampling to achieve a high accuracy. The thickness of the bilayers and the hydrophobic core thickness and the distance between heads increase linearly with the length of the tails, which corresponds to the increased molecular volume of the lipids. The compressibility constant and area per lipid decrease with the length of the tails. This is attributed to the increase of the TT contact energy per lipid with increased number of T beads. To analyze such behavior, we use one of the advantages of the SCMF theory, namely the direct access to components of the total free energy at equilibrium. Figure \ref{fig:energy} shows the dependence of the dominant terms in the total free energy: the energy of TT contacts and the entropy of lipids as a function of the lipid tails length. Other terms to the free energy, the intra-molecular internal energy and the energy of head-solvent interactions are small and almost constant, thus not shown. Main contribution to the free energy per lipid is the energy of TT contacts, which increases with the number of T beads in the tails, making the lipid more hydrophobic and thus leading to closer packing in the bilayer core region.



\begin{figure}[!ht]
\centering
\includegraphics[width=8.5cm]{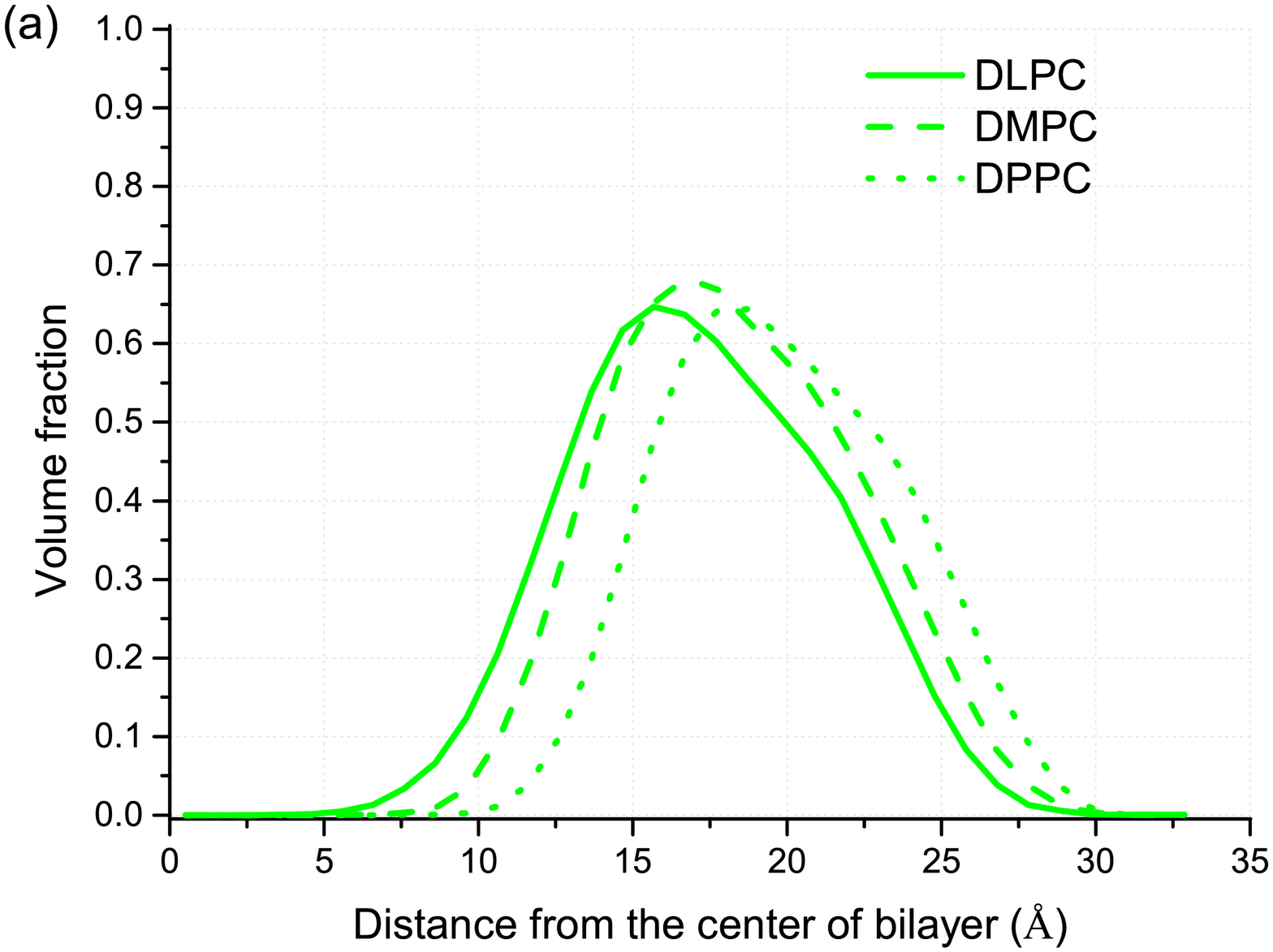}
\includegraphics[width=8.5cm]{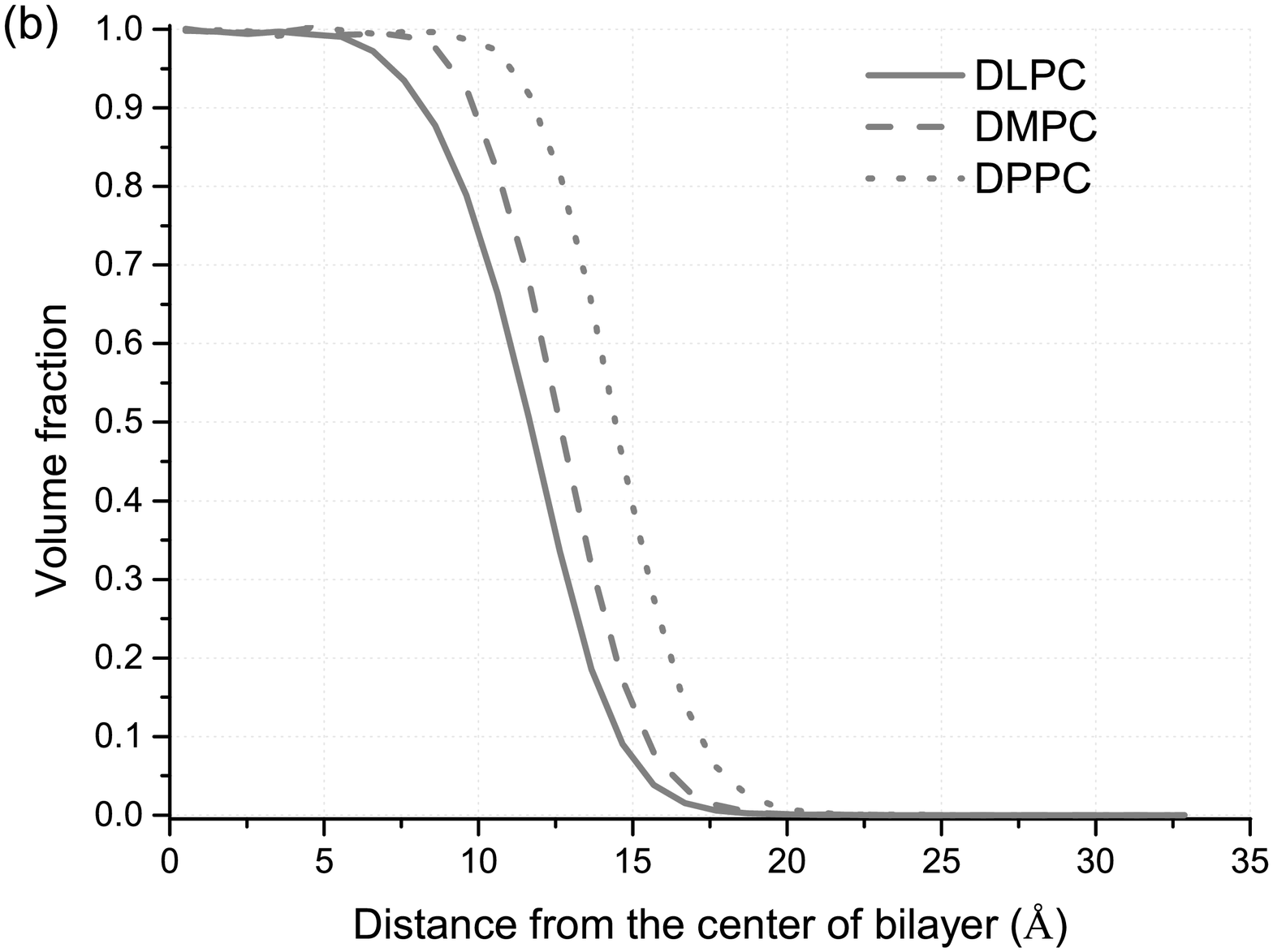}
\includegraphics[width=8.5cm]{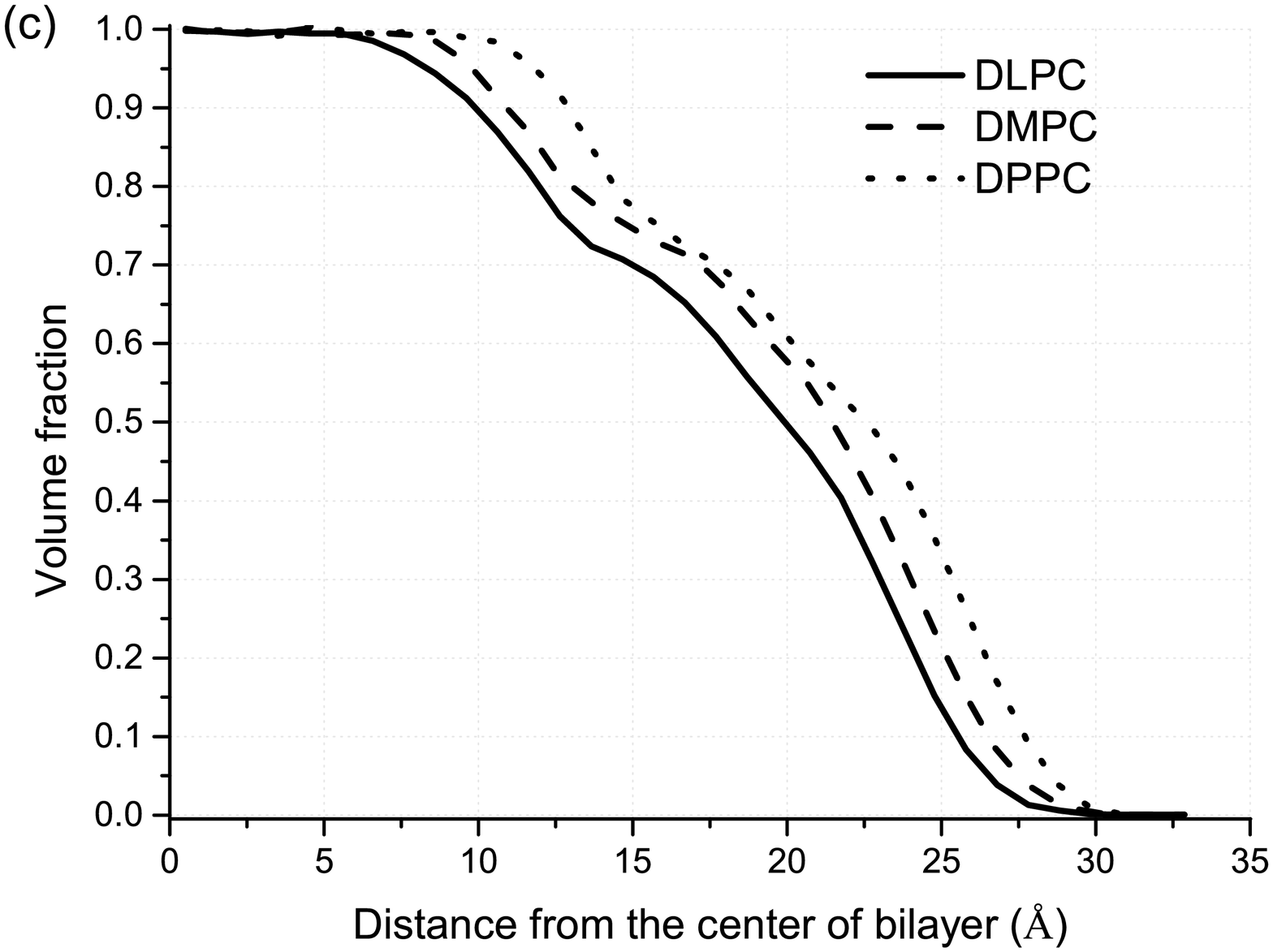}
\caption{Volume fraction profiles of (a) hydrophilic bead H, (b) hydrophobic bead T and c) total volume fraction profile of three saturated lipids given by SCMF model.}
\label{fig:3}
\end{figure}

The average volume fraction profile of the equilibrium
lipid bilayer is shown in Figure~\ref{fig:3}. It shows the total volume fraction in the bilayer as well as the distribution of heads and tails in the bilayer for three saturated lipids. Increased length of the tail leads to the increase of the thickness and increase of the head-to-head distance as reflected in Table \ref{tab:2}.

\begin{figure}[!htb]
\centering
\includegraphics[width=8.5cm]{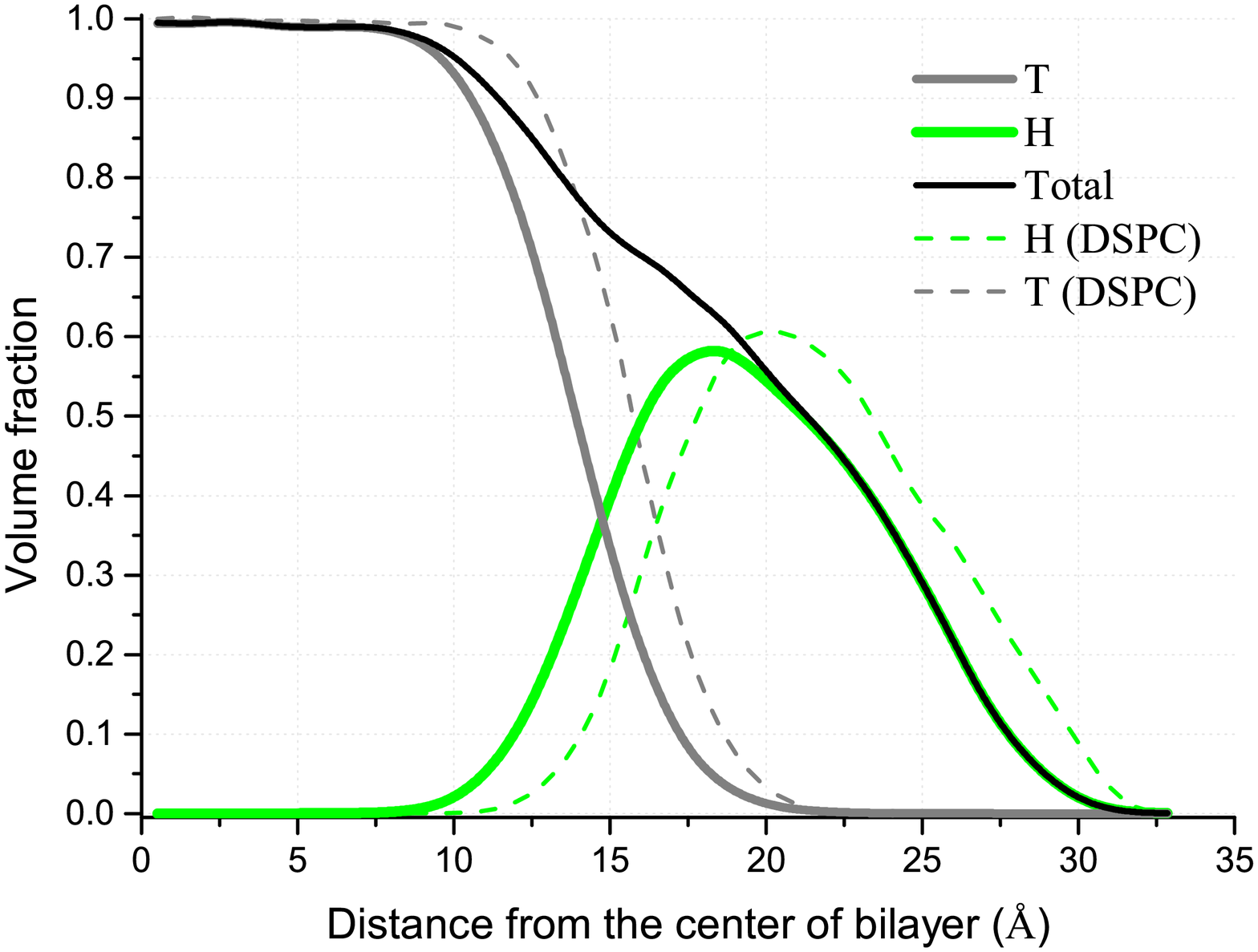}
\caption{Average volume fraction profile of unsaturated DOPC lipid bilayer (solid) as compared with the saturated DSPC lipid bilayer of similar structure (dashed). T represents hydrophobic and H hydrophilic heads.}
\label{fig:5}
\end{figure}

\subsection{Unsaturated lipids}

Our model of unsaturated DOPC and POPC lipids is based on the structure of saturated DPPC or DSPC lipids which have similar tail length and molecular volume, Figure \ref{fig:saturated}. The only difference between saturated and unsaturated lipids is the hydrogen bonds in the middle of the tails. One possibility is to model double bounds in the tails is to fix the angle as shown in Figure \ref{fig:angle} and keep all other parameters of saturated lipids model shown in Table \ref{tab:1}. Such model with no fitting or additional parameters gives reasonable estimates for the structure of the bilayer, thickness of the membrane and hydrophobic core, molecular volume and gives reasonable compressibility constant (red circles in Figure \ref{fig:2}). However, the resulting area per lipid is significantly lower experimental values. This can be explained by underestimated crowding effect induced by disordered packing of tails with fixed angles. But this packing effect relies on strong correlations between neighboring lipid tails and this effect definitely goes beyond mean field and hence is not present in SCMF theory. In addition, these tails with fixed angle in the middle of the tail is difficult to align in parallel arrays, thus, impeding unsaturated lipids from phase transition to gel phase, observed for saturated lipids of the same structure and same temperature.
\begin{figure}[!htb]
\centering
\includegraphics[width=8.5cm]{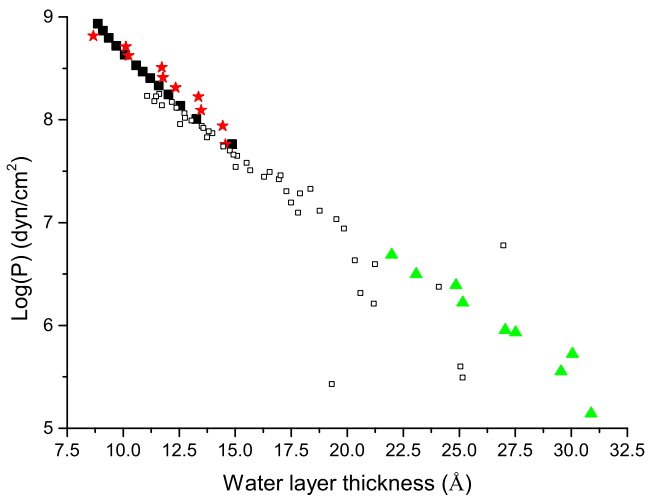}
\caption{Disjoining pressure between two compressed POPC bilayers as a function of water layer thickness $D_{w}$ in a logarithmic scale. Solid and open squares denote experimental and simulation data by \citeauthor{Smirnova2013}\cite{Smirnova2013}, solid triangles are experimental data by \citeauthor{varma_comparison_1989}\cite{varma_comparison_1989}, solid stars are our simulation results with SCMF model.}
\label{fig:7}
\end{figure}
Fixed angle induces the distortion in the conformations of lipids which results in slightly increased average excluded volume of conformations. According to our numerical estimations based on averaging of the excluded volume of each generated conformation, the increase of the excluded volume is about 5.1\% for DOPC and 2.6\% for POPC. This effect was indirectly incorporated into the model of unsaturated lipid by increasing the volume of the beads by this amount. This increase of the volume of the beads has led to significantly better correspondence with experimental results as shown in Figure \ref{fig:2} and Table \ref{tab:2}. The consequences on the lipid bilayer structure induced by the replacement of DSPC by DOPC lipid is depicted in Figure \ref{fig:5}. The thickness of the bilayer of unsaturated lipid is lower than that of saturated lipid because the number of lipids in the equilibrium is smaller.

\section{Bilayers under compression}

General model for saturated phospholipids is shown to be successful in description of equilibrium thermodynamic properties of single unconstrained bilayers. In order to test this model further, we consider compressed bilayers.

Double bilayer systems can be formed in two distinct experimental situations: (i) dehydration of the water layer between the bilayers, when the distance between the bilayers is controlled by hydration level (the number of water molecules per lipid)\cite{schneck_hydration_2012,Smirnova2013} and (ii) mechanical compression of two bilayers\cite{Smirnova2013}, for example, squeezed between two parallel walls; in this case the control parameter is the distance between the centers of mass of the bilayers.

\begin{figure*}[!htb]
\centering
\includegraphics[width=12cm]{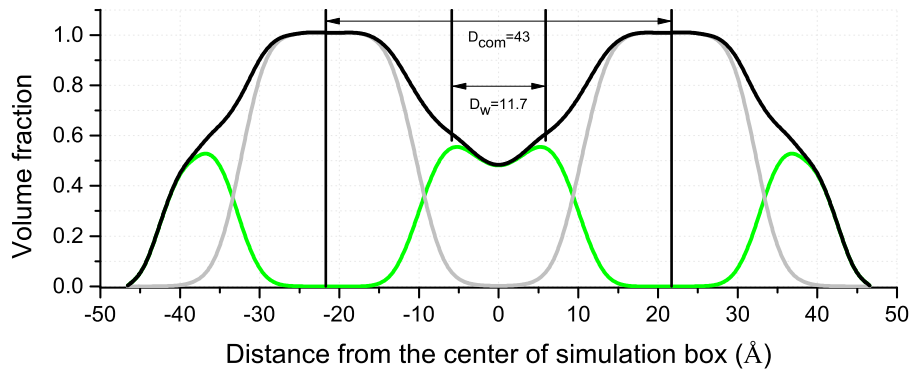}
\caption{Typical volume fraction profile of two compressed bilayers between two walls for POPC lipid bilayer. Green line corresponds to heads, grey line corresponds to tails and black line is the total volume fraction.}
\label{fig:6}
\end{figure*}

We choose two POPC lipid bilayers placed between non-interacting walls in order to compare with the existing experiment and simulation data. The height of the box is decreased and the number of lipids in the box is adjusted to find the minimum of the free energy. The number of lipids corresponding to the minimum of the free energy is then corresponds to equilibrium area per lipid. This is similar in spirit to simulations in Grand canonical ensemble, while mechanical and thermodynamic properties of the bilayers are calculated the same way as discussed in previous section. Figure \ref{fig:6} shows a typical volume fraction profile of two compressed bilayers. In addition, another important property of the bilayer can be calculated, the disjoining pressure at a certain water layer thickness $D_{w}$ which is given by\cite{Smirnova2013}
\begin{equation}
P=\frac{2K_{C}}{D_{com}-D_{w}}(\frac{A_{D_{com}}}{A_{0}}-1)
\label{Pres}
\end{equation}%
where $D_{com}$ is the distance between the centers of the bilayers, $K_{C}$ is the compressibility modulus\cite{pogodin_coarse-grained_2010} and $A_{0}$ is the area per lipid of a unperturbed POPC bilayer at equilibrium. In turn, $D_{com}-D_{w}$ denotes the thickness of single 

bilayer.\cite{Smirnova2013} It is equal to the distance between heads of single bilayer at equilibrium, $D_{hh}$. Since our method can directly measure both $D_{hh}$ and $D_{w}$, they are directly obtained from the volume fraction profile as will be mentioned later. $A_{D_{com}}$ is the area per lipid at a given membrane thickness ($D_{hh}$). The initial height of box is 120 $\AA$ and decreased to 60 $\AA$ by step 5 $\AA$. The number of lipids, free energy per lipid and area per lipid at equilibrium state of compressed bilayers are changed with decreasing volume of the box.


Figure~\ref{fig:7} shows a comparison of disjoining pressure corresponding to water layer thickness, collecting data from both experiments and simulations. The experimental results\cite{varma_comparison_1989} are calculated from osmotic pressure at room temperature. \cite{leneveu_measurement_1976,mccabe_expression_1992} Open square represents the simulation data obtained from compressed double bilayer system using Eq.(\ref{Pres}). Solid star represents the simulation results with SCMF theory for POPC and calculated from Eq.(\ref{Pres}). Here the value of $K_C$ is taken from the Table~\ref{tab:2} as 237 $dyn/cm$. It is important to note that the way to define the thicknesses $D_{w}$ and $D_{hh}$ has a great impact on the result. Several methods of definition have been explored in the literature in both experiment and
simulation. Most common definition comes from X-Ray scattering is the distance between the maxima in the electron density profile\cite{kuvcerka2006structure} which it often related to phosphate peaks. Here we use the same method to define $D_{hh}$ in order to compare with experiment results. Since our model do not distinguish phosphate groups but has only a general hydrophilic head group, $D_{hh}$ is defined as the head-to-head distance of a single unperturbed bilayer, $D_H$ in Table \ref{tab:2}. The water interface of $D_{w}$  is defined at the half density of solvent in our simulation. Figure~\ref{fig:7} shows a good agreement with both experiment and simulation of our  model in a compressed two-bilayer system.

\begin{figure}[!htb]
\centering
\includegraphics[width=8.5cm]{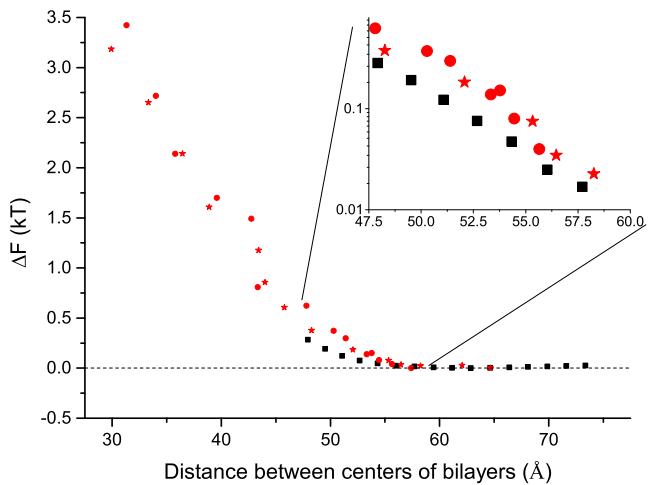}
\caption{Free energy cost $\Delta F$ versus the center of mass between 2 bilayers. Solid squares represents the MD simulation data obtained by \citeauthor{Smirnova2013}\cite{Smirnova2013} for POPC lipid bilayer with MARTINI model, stars are SCMF simulation results for POPC lipid bilayers and solid circles are SCMF simulation for DMPC lipid bilayers.}
\label{fig:8}
\end{figure}

\begin{figure}[!htb]
\centering \includegraphics[width=8.5cm]{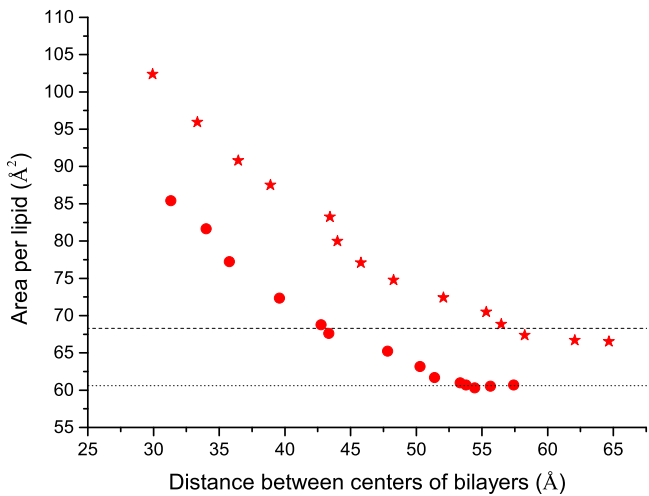}
\caption{Area per lipid versus centers of mass between two bilayers $D_{com}$, solid circle (DMPC) and star (POPC) represent the simulation results with SCMF theory. Dashed (DMPC) and dotted (POPC) lines denote the experimental value for unperturbed single bilayers for reference at infinite distances.}
\label{fig:area}
\end{figure}

The free energy cost of the equilibrium state of compressed double bilayers as a function of the distance between the centers of mass of the bilayers is shown in Figure~\ref{fig:8}. The free energy difference per lipid is calculated as a function of the free energy of unperturbed single bilayer $\Delta F=(F-F_{eq})$. We compare our simulation results (red stars and circles) with a similar double bilayer system within MD simulations using MARTINI model\cite{Smirnova2013} (black squares). In highly compressed state, $D_{com} < 45$ $\AA$, we find linear dependence between the free energy and the distance between centers of bilayers. For small compressions, $D_{com} > 45$ $\AA$, our simulations show a very good agreement with molecular dynamics simulation.

Figure~\ref{fig:area} shows the area per lipid as a distance between the centers of bilayers calculated with SCMF theory. Dashed lines correspond to equilibrium areas per lipid in unperturbed single bilayers of DMPC and POPC membranes, which corresponds to two bilayers that do not feel each other. With increasing distance between bilayers the area per lipid decreases with in a similar way for saturated DMPC and unsaturated POPC lipid bilayers.

\section{Conclusion}

We propose a general model for saturated, DCPC, DLPC, DMPC, DPPC,
DSPC and unsaturated POPC and DOPC phospholipids which differs only in hydrocarbon chains. The lipid molecule is represented by two hydrophilic beads which are the same for all studied lipids and 6-10 beads in the tail and correspond to different number of carbons. Essential
equilibrium properties of the phospholipid bilayer such as compressibility
constant, volume fraction, and the area per lipid can be obtained with good
precision and in accordance with experimental data. This general model is able to
describe most of equilibrium properties of phospholipid bilayers such as the
thickness of the bilayer and the hydrophobic core, position of hydrophilic
and hydrophobic groups in the bilayer, the mechanical properties of the
bilayer and the corresponding compressibility constant. The performance of
the model was tested in a two-bilayer system, where non-adsorbing membrane
is compressed and the force -- distance profile is measured. The model gives
results, which are in good agreement with experimental data as well as with
molecular dynamics simulations. The proposed model can further be used for modeling thermodynamic and mechanical properties of mixtures of lipids and interactions with nano-objects.

\acknowledgments{Authors are grateful to Dr. Yuliya G. Smirnova for useful comments and discussions.}

\pagebreak

\begin{thebibliography}{32}%
\makeatletter
\providecommand \@ifxundefined [1]{%
 \@ifx{#1\undefined}
}%
\providecommand \@ifnum [1]{%
 \ifnum #1\expandafter \@firstoftwo
 \else \expandafter \@secondoftwo
 \fi
}%
\providecommand \@ifx [1]{%
 \ifx #1\expandafter \@firstoftwo
 \else \expandafter \@secondoftwo
 \fi
}%
\providecommand \natexlab [1]{#1}%
\providecommand \enquote  [1]{``#1''}%
\providecommand \bibnamefont  [1]{#1}%
\providecommand \bibfnamefont [1]{#1}%
\providecommand \citenamefont [1]{#1}%
\providecommand \href@noop [0]{\@secondoftwo}%
\providecommand \href [0]{\begingroup \@sanitize@url \@href}%
\providecommand \@href[1]{\@@startlink{#1}\@@href}%
\providecommand \@@href[1]{\endgroup#1\@@endlink}%
\providecommand \@sanitize@url [0]{\catcode `\\12\catcode `\$12\catcode
  `\&12\catcode `\#12\catcode `\^12\catcode `\_12\catcode `\%12\relax}%
\providecommand \@@startlink[1]{}%
\providecommand \@@endlink[0]{}%
\providecommand \url  [0]{\begingroup\@sanitize@url \@url }%
\providecommand \@url [1]{\endgroup\@href {#1}{\urlprefix }}%
\providecommand \urlprefix  [0]{URL }%
\providecommand \Eprint [0]{\href }%
\providecommand \doibase [0]{http://dx.doi.org/}%
\providecommand \selectlanguage [0]{\@gobble}%
\providecommand \bibinfo  [0]{\@secondoftwo}%
\providecommand \bibfield  [0]{\@secondoftwo}%
\providecommand \translation [1]{[#1]}%
\providecommand \BibitemOpen [0]{}%
\providecommand \bibitemStop [0]{}%
\providecommand \bibitemNoStop [0]{.\EOS\space}%
\providecommand \EOS [0]{\spacefactor3000\relax}%
\providecommand \BibitemShut  [1]{\csname bibitem#1\endcsname}%
\let\auto@bib@innerbib\@empty
\bibitem [{\citenamefont {Alberts}(2008)}]{alberts_molecular_2008}%
  \BibitemOpen
  \bibfield  {author} {\bibinfo {author} {\bibfnamefont {B.}~\bibnamefont
  {Alberts}},\ }\href@noop {} {\emph {\bibinfo {title} {Molecular Biology of
  the Cell}}},\ \bibinfo {edition} {5th}\ ed.\ (\bibinfo  {publisher} {Garland
  Science},\ \bibinfo {address} {New York},\ \bibinfo {year}
  {2008})\BibitemShut {NoStop}%
\bibitem [{\citenamefont {Yeagle}(2005)}]{yeagle_structure_2005}%
  \BibitemOpen
  \bibfield  {author} {\bibinfo {author} {\bibfnamefont {P.~L.}\ \bibnamefont
  {Yeagle}},\ }\href@noop {} {\emph {\bibinfo {title} {The Structure of
  Biological Membranes}}},\ \bibinfo {edition} {2nd}\ ed.\ (\bibinfo
  {publisher} {{CRC} Press},\ \bibinfo {year} {2005})\BibitemShut {NoStop}%
\bibitem [{\citenamefont {Dopico}\ and\ \citenamefont
  {Tigyi}()}]{dopico2007glance}%
  \BibitemOpen
  \bibfield  {author} {\bibinfo {author} {\bibfnamefont {A.~M.}\ \bibnamefont
  {Dopico}}\ and\ \bibinfo {author} {\bibfnamefont {G.~J.}\ \bibnamefont
  {Tigyi}},\ }\href@noop {} {\bibfield  {journal} {\bibinfo  {journal} {Methods
  Mol. Biol.}\ }\textbf {\bibinfo {volume} {400}},\ \bibinfo {pages}
  {1}}\BibitemShut {NoStop}%
\bibitem [{\citenamefont {Halliwell}\ and\ \citenamefont
  {Gutteridge}(1990)}]{halliwell_role_1990}%
  \BibitemOpen
  \bibfield  {author} {\bibinfo {author} {\bibfnamefont {B.}~\bibnamefont
  {Halliwell}}\ and\ \bibinfo {author} {\bibfnamefont {J.~M.}\ \bibnamefont
  {Gutteridge}},\ }in\ \href
  {http://linkinghub.elsevier.com/retrieve/pii/007668799086093B} {\emph
  {\bibinfo {booktitle} {Methods in Enzymology}}},\ Vol.\ \bibinfo {volume}
  {186}\ (\bibinfo  {publisher} {Elsevier},\ \bibinfo {year} {1990})\ pp.\
  \bibinfo {pages} {1--85}\BibitemShut {NoStop}%
\bibitem [{\citenamefont {M\"uller}\ \emph {et~al.}(2003)\citenamefont
  {M\"uller}, \citenamefont {Katsov},\ and\ \citenamefont
  {Schick}}]{muller_coarse-grained_2003}%
  \BibitemOpen
  \bibfield  {author} {\bibinfo {author} {\bibfnamefont {M.}~\bibnamefont
  {M\"uller}}, \bibinfo {author} {\bibfnamefont {K.}~\bibnamefont {Katsov}}, \
  and\ \bibinfo {author} {\bibfnamefont {M.}~\bibnamefont {Schick}},\
  }\href@noop {} {\bibfield  {journal} {\bibinfo  {journal} {J. Polym. Sci. B:
  Polym. Phys.}\ }\textbf {\bibinfo {volume} {41}},\ \bibinfo {pages}
  {1441??450} (\bibinfo {year} {2003})}\BibitemShut {NoStop}%
\bibitem [{\citenamefont {M\"uller}\ \emph {et~al.}(2006)\citenamefont
  {M\"uller}, \citenamefont {Katsov},\ and\ \citenamefont
  {Schick}}]{muller_biological_2006}%
  \BibitemOpen
  \bibfield  {author} {\bibinfo {author} {\bibfnamefont {M.}~\bibnamefont
  {M\"uller}}, \bibinfo {author} {\bibfnamefont {K.}~\bibnamefont {Katsov}}, \
  and\ \bibinfo {author} {\bibfnamefont {M.}~\bibnamefont {Schick}},\ }\href
  {\doibase 10.1016/j.physrep.2006.08.003} {\bibfield  {journal} {\bibinfo
  {journal} {Phys Rep}\ }\textbf {\bibinfo {volume} {434}},\ \bibinfo {pages}
  {113} (\bibinfo {year} {2006})}\BibitemShut {NoStop}%
\bibitem [{\citenamefont {Venturoli}\ \emph {et~al.}(2006)\citenamefont
  {Venturoli}, \citenamefont {Sperotto}, \citenamefont {Kranenburg},\ and\
  \citenamefont {Smit}}]{venturoli_mesoscopic_2006}%
  \BibitemOpen
  \bibfield  {author} {\bibinfo {author} {\bibfnamefont {M.}~\bibnamefont
  {Venturoli}}, \bibinfo {author} {\bibfnamefont {M.~M.}\ \bibnamefont
  {Sperotto}}, \bibinfo {author} {\bibfnamefont {M.}~\bibnamefont
  {Kranenburg}}, \ and\ \bibinfo {author} {\bibfnamefont {B.}~\bibnamefont
  {Smit}},\ }\href@noop {} {\bibfield  {journal} {\bibinfo  {journal} {Phys
  Rep}\ }\textbf {\bibinfo {volume} {437}},\ \bibinfo {pages} {1} (\bibinfo
  {year} {2006})}\BibitemShut {NoStop}%
\bibitem [{\citenamefont {Lyubartsev}(2005)}]{lyubartsev_multiscale_2005}%
  \BibitemOpen
  \bibfield  {author} {\bibinfo {author} {\bibfnamefont {A.~P.}\ \bibnamefont
  {Lyubartsev}},\ }\href@noop {} {\bibfield  {journal} {\bibinfo  {journal}
  {Eur Biophys J}\ }\textbf {\bibinfo {volume} {35}},\ \bibinfo {pages} {53}
  (\bibinfo {year} {2005})}\BibitemShut {NoStop}%
\bibitem [{\citenamefont {Pogodin}\ and\ \citenamefont
  {Baulin}(2010)}]{pogodin_coarse-grained_2010}%
  \BibitemOpen
  \bibfield  {author} {\bibinfo {author} {\bibfnamefont {S.}~\bibnamefont
  {Pogodin}}\ and\ \bibinfo {author} {\bibfnamefont {V.~A.}\ \bibnamefont
  {Baulin}},\ }\href {\doibase 10.1039/B927437E} {\bibfield  {journal}
  {\bibinfo  {journal} {Soft Matter}\ }\textbf {\bibinfo {volume} {6}},\
  \bibinfo {pages} {2216} (\bibinfo {year} {2010})}\BibitemShut {NoStop}%
\bibitem [{\citenamefont {Ben-Shaul}\ \emph {et~al.}(1985)\citenamefont
  {Ben-Shaul}, \citenamefont {Szleifer},\ and\ \citenamefont
  {Gelbart}}]{ben-shaul_chain_1985}%
  \BibitemOpen
  \bibfield  {author} {\bibinfo {author} {\bibfnamefont {A.}~\bibnamefont
  {Ben-Shaul}}, \bibinfo {author} {\bibfnamefont {I.}~\bibnamefont {Szleifer}},
  \ and\ \bibinfo {author} {\bibfnamefont {W.~M.}\ \bibnamefont {Gelbart}},\
  }\href@noop {} {\bibfield  {journal} {\bibinfo  {journal} {J. Chem. Phys.}\
  }\textbf {\bibinfo {volume} {83}},\ \bibinfo {pages} {3597} (\bibinfo {year}
  {1985})}\BibitemShut {NoStop}%
\bibitem [{\citenamefont {Szleifer}\ \emph {et~al.}(1985)\citenamefont
  {Szleifer}, \citenamefont {Ben-Shaul},\ and\ \citenamefont
  {Gelbart}}]{szleifer_chain_1985}%
  \BibitemOpen
  \bibfield  {author} {\bibinfo {author} {\bibfnamefont {I.}~\bibnamefont
  {Szleifer}}, \bibinfo {author} {\bibfnamefont {A.}~\bibnamefont {Ben-Shaul}},
  \ and\ \bibinfo {author} {\bibfnamefont {W.~M.}\ \bibnamefont {Gelbart}},\
  }\href {\doibase 10.1063/1.449167} {\bibfield  {journal} {\bibinfo  {journal}
  {J. Chem. Phys.}\ }\textbf {\bibinfo {volume} {83}},\ \bibinfo {pages} {3612}
  (\bibinfo {year} {1985})}\BibitemShut {NoStop}%
\bibitem [{\citenamefont {Ben-Shaul}\ \emph {et~al.}(1986)\citenamefont
  {Ben-Shaul}, \citenamefont {Szleifer},\ and\ \citenamefont
  {Gelbart}}]{ben-shaul_chain_1986}%
  \BibitemOpen
  \bibfield  {author} {\bibinfo {author} {\bibfnamefont {A.}~\bibnamefont
  {Ben-Shaul}}, \bibinfo {author} {\bibfnamefont {I.}~\bibnamefont {Szleifer}},
  \ and\ \bibinfo {author} {\bibfnamefont {W.~M.}\ \bibnamefont {Gelbart}},\
  }\href@noop {} {\bibfield  {journal} {\bibinfo  {journal} {J. Chem. Phys.}\
  }\textbf {\bibinfo {volume} {85}},\ \bibinfo {pages} {5345} (\bibinfo {year}
  {1986})}\BibitemShut {NoStop}%
\bibitem [{\citenamefont {Al-Anber}\ \emph {et~al.}(2005)\citenamefont
  {Al-Anber}, \citenamefont {Bonet-Avalos},\ and\ \citenamefont
  {Mackie}}]{al-anber_prediction_2005}%
  \BibitemOpen
  \bibfield  {author} {\bibinfo {author} {\bibfnamefont {Z.~A.}\ \bibnamefont
  {Al-Anber}}, \bibinfo {author} {\bibfnamefont {J.}~\bibnamefont
  {Bonet-Avalos}}, \ and\ \bibinfo {author} {\bibfnamefont {A.~D.}\
  \bibnamefont {Mackie}},\ }\href@noop {} {\bibfield  {journal} {\bibinfo
  {journal} {J. Chem. Phys.}\ }\textbf {\bibinfo {volume} {122}},\ \bibinfo
  {pages} {104910} (\bibinfo {year} {2005})}\BibitemShut {NoStop}%
\bibitem [{\citenamefont {Gezae~Daful}\ \emph {et~al.}(2011)\citenamefont
  {Gezae~Daful}, \citenamefont {Baulin}, \citenamefont {Bonet~Avalos},\ and\
  \citenamefont {Mackie}}]{gezae_daful_accurate_2011}%
  \BibitemOpen
  \bibfield  {author} {\bibinfo {author} {\bibfnamefont {A.}~\bibnamefont
  {Gezae~Daful}}, \bibinfo {author} {\bibfnamefont {V.~A.}\ \bibnamefont
  {Baulin}}, \bibinfo {author} {\bibfnamefont {J.}~\bibnamefont
  {Bonet~Avalos}}, \ and\ \bibinfo {author} {\bibfnamefont {A.~D.}\
  \bibnamefont {Mackie}},\ }\href {\doibase 10.1021/jp1102302} {\bibfield
  {journal} {\bibinfo  {journal} {J. Phys. Chem. B}\ }\textbf {\bibinfo
  {volume} {115}},\ \bibinfo {pages} {3434} (\bibinfo {year}
  {2011})}\BibitemShut {NoStop}%
\bibitem [{\citenamefont {Daoulas}\ and\ \citenamefont
  {Muller}(2006)}]{daoulas_single_2006}%
  \BibitemOpen
  \bibfield  {author} {\bibinfo {author} {\bibfnamefont {K.~C.}\ \bibnamefont
  {Daoulas}}\ and\ \bibinfo {author} {\bibfnamefont {M.}~\bibnamefont
  {Muller}},\ }\href@noop {} {\bibfield  {journal} {\bibinfo  {journal} {J.
  Chem. Phys.}\ }\textbf {\bibinfo {volume} {125}},\ \bibinfo {pages} {184904}
  (\bibinfo {year} {2006})}\BibitemShut {NoStop}%
\bibitem [{\citenamefont {Mouritsen}(2005)}]{mouritsen2005life}%
  \BibitemOpen
  \bibfield  {author} {\bibinfo {author} {\bibfnamefont {O.~G.}\ \bibnamefont
  {Mouritsen}},\ }\href@noop {} {\emph {\bibinfo {title} {Life-as a matter of
  fat: the emerging science of lipidomics}}}\ (\bibinfo  {publisher}
  {Springer},\ \bibinfo {year} {2005})\BibitemShut {NoStop}%
\bibitem [{\citenamefont {Smirnova}\ \emph {et~al.}(2013)\citenamefont
  {Smirnova}, \citenamefont {Aeffner}, \citenamefont {Risselada}, \citenamefont
  {Salditt}, \citenamefont {Marrink}, \citenamefont {Mueller},\ and\
  \citenamefont {Knecht}}]{Smirnova2013}%
  \BibitemOpen
  \bibfield  {author} {\bibinfo {author} {\bibfnamefont {Y.}~\bibnamefont
  {Smirnova}}, \bibinfo {author} {\bibfnamefont {S.}~\bibnamefont {Aeffner}},
  \bibinfo {author} {\bibfnamefont {J.}~\bibnamefont {Risselada}}, \bibinfo
  {author} {\bibfnamefont {T.}~\bibnamefont {Salditt}}, \bibinfo {author}
  {\bibfnamefont {S.~J.}\ \bibnamefont {Marrink}}, \bibinfo {author}
  {\bibfnamefont {M.}~\bibnamefont {Mueller}}, \ and\ \bibinfo {author}
  {\bibfnamefont {V.}~\bibnamefont {Knecht}},\ }\href {\doibase
  10.1039/c3sm51771c} {\bibfield  {journal} {\bibinfo  {journal} {Soft Matter}\
  }\textbf {\bibinfo {volume} {9}},\ \bibinfo {pages} {10705} (\bibinfo {year}
  {2013})}\BibitemShut {NoStop}%
\bibitem [{\citenamefont {Pogodin}\ and\ \citenamefont
  {Baulin}(2011)}]{pogodin_equilibrium_2011}%
  \BibitemOpen
  \bibfield  {author} {\bibinfo {author} {\bibfnamefont {S.}~\bibnamefont
  {Pogodin}}\ and\ \bibinfo {author} {\bibfnamefont {V.}~\bibnamefont
  {Baulin}},\ }\href@noop {} {\bibfield  {journal} {\bibinfo  {journal} {Curr
  Nanosci}\ }\textbf {\bibinfo {volume} {7}},\ \bibinfo {pages} {721} (\bibinfo
  {year} {2011})}\BibitemShut {NoStop}%
\bibitem [{fle(1993)}]{fleer_polymers_1993}%
  \BibitemOpen
  \href@noop {} {\emph {\bibinfo {title} {Polymers at interfaces}}},\ \bibinfo
  {edition} {1st}\ ed.\ (\bibinfo  {publisher} {Chapman \& Hall},\ \bibinfo
  {address} {London ; New York},\ \bibinfo {year} {1993})\BibitemShut {NoStop}%
\bibitem [{\citenamefont {Thompson}\ \emph {et~al.}(2012)\citenamefont
  {Thompson}, \citenamefont {Jebb},\ and\ \citenamefont
  {Wen}}]{thompson_benchmarking_2012}%
  \BibitemOpen
  \bibfield  {author} {\bibinfo {author} {\bibfnamefont {R.~B.}\ \bibnamefont
  {Thompson}}, \bibinfo {author} {\bibfnamefont {T.}~\bibnamefont {Jebb}}, \
  and\ \bibinfo {author} {\bibfnamefont {Y.}~\bibnamefont {Wen}},\ }\href
  {\doibase 10.1039/c2sm26352a} {\bibfield  {journal} {\bibinfo  {journal}
  {Soft Matter}\ }\textbf {\bibinfo {volume} {8}},\ \bibinfo {pages} {9877}
  (\bibinfo {year} {2012})}\BibitemShut {NoStop}%
\bibitem [{\citenamefont {Rosenbluth}\ and\ \citenamefont
  {Rosenbluth}(1955)}]{rosenbluth_monte_1955}%
  \BibitemOpen
  \bibfield  {author} {\bibinfo {author} {\bibfnamefont {M.~N.}\ \bibnamefont
  {Rosenbluth}}\ and\ \bibinfo {author} {\bibfnamefont {A.~W.}\ \bibnamefont
  {Rosenbluth}},\ }\href@noop {} {\bibfield  {journal} {\bibinfo  {journal} {J.
  Chem. Phys.}\ }\textbf {\bibinfo {volume} {23}},\ \bibinfo {pages} {356}
  (\bibinfo {year} {1955})}\BibitemShut {NoStop}%
\bibitem [{\citenamefont {Ku{\v{c}}erka}\ \emph {et~al.}(2005)\citenamefont
  {Ku{\v{c}}erka}, \citenamefont {Liu}, \citenamefont {Chu}, \citenamefont
  {Petrache}, \citenamefont {Tristram-Nagle},\ and\ \citenamefont
  {Nagle}}]{kucerka_structure_2005}%
  \BibitemOpen
  \bibfield  {author} {\bibinfo {author} {\bibfnamefont {N.}~\bibnamefont
  {Ku{\v{c}}erka}}, \bibinfo {author} {\bibfnamefont {Y.}~\bibnamefont {Liu}},
  \bibinfo {author} {\bibfnamefont {N.}~\bibnamefont {Chu}}, \bibinfo {author}
  {\bibfnamefont {H.~I.}\ \bibnamefont {Petrache}}, \bibinfo {author}
  {\bibfnamefont {S.}~\bibnamefont {Tristram-Nagle}}, \ and\ \bibinfo {author}
  {\bibfnamefont {J.~F.}\ \bibnamefont {Nagle}},\ }\href {\doibase
  10.1529/biophysj.104.056606} {\bibfield  {journal} {\bibinfo  {journal}
  {Biophys. J.}\ }\textbf {\bibinfo {volume} {88}},\ \bibinfo {pages} {2626}
  (\bibinfo {year} {2005})}\BibitemShut {NoStop}%
\bibitem [{\citenamefont {Ku{\v{c}}erka}\ \emph {et~al.}(2006)\citenamefont
  {Ku{\v{c}}erka}, \citenamefont {Tristram-Nagle},\ and\ \citenamefont
  {Nagle}}]{kuvcerka2006structure}%
  \BibitemOpen
  \bibfield  {author} {\bibinfo {author} {\bibfnamefont {N.}~\bibnamefont
  {Ku{\v{c}}erka}}, \bibinfo {author} {\bibfnamefont {S.}~\bibnamefont
  {Tristram-Nagle}}, \ and\ \bibinfo {author} {\bibfnamefont {J.~F.}\
  \bibnamefont {Nagle}},\ }\href@noop {} {\bibfield  {journal} {\bibinfo
  {journal} {J. Membr. Biol.}\ }\textbf {\bibinfo {volume} {208}},\ \bibinfo
  {pages} {193} (\bibinfo {year} {2006})}\BibitemShut {NoStop}%
\bibitem [{\citenamefont {Nagle}\ \emph {et~al.}(1996)\citenamefont {Nagle},
  \citenamefont {Zhang}, \citenamefont {Tristram-Nagle}, \citenamefont {Sun},
  \citenamefont {Petrache},\ and\ \citenamefont {Suter}}]{nagle1996x}%
  \BibitemOpen
  \bibfield  {author} {\bibinfo {author} {\bibfnamefont {J.~F.}\ \bibnamefont
  {Nagle}}, \bibinfo {author} {\bibfnamefont {R.}~\bibnamefont {Zhang}},
  \bibinfo {author} {\bibfnamefont {S.}~\bibnamefont {Tristram-Nagle}},
  \bibinfo {author} {\bibfnamefont {W.}~\bibnamefont {Sun}}, \bibinfo {author}
  {\bibfnamefont {H.~I.}\ \bibnamefont {Petrache}}, \ and\ \bibinfo {author}
  {\bibfnamefont {R.~M.}\ \bibnamefont {Suter}},\ }\href@noop {} {\bibfield
  {journal} {\bibinfo  {journal} {Biophys. J.}\ }\textbf {\bibinfo {volume}
  {70}},\ \bibinfo {pages} {1419} (\bibinfo {year} {1996})}\BibitemShut
  {NoStop}%
\bibitem [{\citenamefont {Feig}(2008)}]{feig2008implicit}%
  \BibitemOpen
  \bibfield  {author} {\bibinfo {author} {\bibfnamefont {M.}~\bibnamefont
  {Feig}},\ }in\ \href@noop {} {\emph {\bibinfo {booktitle} {Molecular Modeling
  of Proteins}}}\ (\bibinfo  {publisher} {Springer},\ \bibinfo {year} {2008})\
  pp.\ \bibinfo {pages} {181--196}\BibitemShut {NoStop}%
\bibitem [{\citenamefont {Mathai}\ \emph {et~al.}(2008)\citenamefont {Mathai},
  \citenamefont {Tristram-Nagle}, \citenamefont {Nagle},\ and\ \citenamefont
  {Zeidel}}]{mathai_structural_2008}%
  \BibitemOpen
  \bibfield  {author} {\bibinfo {author} {\bibfnamefont {J.~C.}\ \bibnamefont
  {Mathai}}, \bibinfo {author} {\bibfnamefont {S.}~\bibnamefont
  {Tristram-Nagle}}, \bibinfo {author} {\bibfnamefont {J.~F.}\ \bibnamefont
  {Nagle}}, \ and\ \bibinfo {author} {\bibfnamefont {M.~L.}\ \bibnamefont
  {Zeidel}},\ }\href {\doibase 10.1085/jgp.200709848} {\bibfield  {journal}
  {\bibinfo  {journal} {J. Gen. Physiol.}\ }\textbf {\bibinfo {volume} {131}},\
  \bibinfo {pages} {69} (\bibinfo {year} {2008})}\BibitemShut {NoStop}%
\bibitem [{\citenamefont {Ku{\v{c}}erka}\ \emph {et~al.}(2011)\citenamefont
  {Ku{\v{c}}erka}, \citenamefont {Nieh},\ and\ \citenamefont
  {Katsaras}}]{kucerka_fluid_2011}%
  \BibitemOpen
  \bibfield  {author} {\bibinfo {author} {\bibfnamefont {N.}~\bibnamefont
  {Ku{\v{c}}erka}}, \bibinfo {author} {\bibfnamefont {M.-P.}\ \bibnamefont
  {Nieh}}, \ and\ \bibinfo {author} {\bibfnamefont {J.}~\bibnamefont
  {Katsaras}},\ }\href {\doibase 10.1016/j.bbamem.2011.07.022} {\bibfield
  {journal} {\bibinfo  {journal} {Biochim. Biophys. Acta}\ }\textbf {\bibinfo
  {volume} {1808}},\ \bibinfo {pages} {2761} (\bibinfo {year}
  {2011})}\BibitemShut {NoStop}%
\bibitem [{\citenamefont {Gauger}\ \emph {et~al.}(2001)\citenamefont {Gauger},
  \citenamefont {Selle}, \citenamefont {Fritzsche},\ and\ \citenamefont
  {Pohle}}]{gauger_chain-length_2001}%
  \BibitemOpen
  \bibfield  {author} {\bibinfo {author} {\bibfnamefont {D.}~\bibnamefont
  {Gauger}}, \bibinfo {author} {\bibfnamefont {C.}~\bibnamefont {Selle}},
  \bibinfo {author} {\bibfnamefont {H.}~\bibnamefont {Fritzsche}}, \ and\
  \bibinfo {author} {\bibfnamefont {W.}~\bibnamefont {Pohle}},\ }\href
  {\doibase 10.1016/S0022-2860(00)00777-8} {\bibfield  {journal} {\bibinfo
  {journal} {J. Mol. Struct.}\ }\textbf {\bibinfo {volume} {565}},\ \bibinfo
  {pages} {25} (\bibinfo {year} {2001})}\BibitemShut {NoStop}%
\bibitem [{\citenamefont {Schneck}\ \emph {et~al.}(2012)\citenamefont
  {Schneck}, \citenamefont {Sedlmeier},\ and\ \citenamefont
  {Netz}}]{schneck_hydration_2012}%
  \BibitemOpen
  \bibfield  {author} {\bibinfo {author} {\bibfnamefont {E.}~\bibnamefont
  {Schneck}}, \bibinfo {author} {\bibfnamefont {F.}~\bibnamefont {Sedlmeier}},
  \ and\ \bibinfo {author} {\bibfnamefont {R.~R.}\ \bibnamefont {Netz}},\
  }\href {\doibase 10.1073/pnas.1205811109} {\bibfield  {journal} {\bibinfo
  {journal} {Proc. Natl. Acad. Sci. U. S. A.}\ }\textbf {\bibinfo {volume}
  {109}},\ \bibinfo {pages} {14405} (\bibinfo {year} {2012})}\BibitemShut
  {NoStop}%
\bibitem [{\citenamefont {Varma}\ \emph {et~al.}(1989)\citenamefont {Varma},
  \citenamefont {Taalbi}, \citenamefont {Collins}, \citenamefont {Tamura-Lis},\
  and\ \citenamefont {Lis}}]{varma_comparison_1989}%
  \BibitemOpen
  \bibfield  {author} {\bibinfo {author} {\bibfnamefont {V.}~\bibnamefont
  {Varma}}, \bibinfo {author} {\bibfnamefont {M.}~\bibnamefont {Taalbi}},
  \bibinfo {author} {\bibfnamefont {J.}~\bibnamefont {Collins}}, \bibinfo
  {author} {\bibfnamefont {W.}~\bibnamefont {Tamura-Lis}}, \ and\ \bibinfo
  {author} {\bibfnamefont {L.}~\bibnamefont {Lis}},\ }\href {\doibase
  10.1016/S0021-9797(89)80052-9} {\bibfield  {journal} {\bibinfo  {journal} {J.
  Colloid Interface Sci.}\ }\textbf {\bibinfo {volume} {133}},\ \bibinfo
  {pages} {426} (\bibinfo {year} {1989})}\BibitemShut {NoStop}%
\bibitem [{\citenamefont {Leneveu}\ \emph {et~al.}(1976)\citenamefont
  {Leneveu}, \citenamefont {Rand},\ and\ \citenamefont
  {Parsegian}}]{leneveu_measurement_1976}%
  \BibitemOpen
  \bibfield  {author} {\bibinfo {author} {\bibfnamefont {D.~M.}\ \bibnamefont
  {Leneveu}}, \bibinfo {author} {\bibfnamefont {R.~P.}\ \bibnamefont {Rand}}, \
  and\ \bibinfo {author} {\bibfnamefont {V.~A.}\ \bibnamefont {Parsegian}},\
  }\href {\doibase 10.1038/259601a0} {\bibfield  {journal} {\bibinfo  {journal}
  {Nature}\ }\textbf {\bibinfo {volume} {259}},\ \bibinfo {pages} {601}
  (\bibinfo {year} {1976})}\BibitemShut {NoStop}%
\bibitem [{\citenamefont {{McCabe}}\ and\ \citenamefont
  {Cole}(1992)}]{mccabe_expression_1992}%
  \BibitemOpen
  \bibfield  {author} {\bibinfo {author} {\bibfnamefont {C.~F.}\ \bibnamefont
  {{McCabe}}}\ and\ \bibinfo {author} {\bibfnamefont {G.~J.}\ \bibnamefont
  {Cole}},\ }\href@noop {} {\bibfield  {journal} {\bibinfo  {journal} {Brain
  Res. Dev. Brain Res.}\ }\textbf {\bibinfo {volume} {70}},\ \bibinfo {pages}
  {9} (\bibinfo {year} {1992})}\BibitemShut {NoStop}%
\end{thebibliography}
%

\end{document}